\documentclass[letterpaper]{article}
\usepackage{aaai20}
\usepackage{times}
\usepackage{helvet}
\usepackage{courier}
\usepackage[hyphens]{url} 
\usepackage{graphicx}
\usepackage{flushend}
\usepackage{booktabs}

\newcommand{\citet}[1]{\citeauthor{#1}~\shortcite{#1}}

\title{Expertise and Dynamics within Crowdsourced Musical Knowledge Curation: \\
A Case Study of the Genius Platform}

\author{Derek Lim\\
Cornell University\\
dl772@cornell.edu
\And
Austin R. Benson\\
Cornell University\\
arb@cs.cornell.edu\\
}

\usepackage{amsmath, amssymb, amsthm, bbm, mathtools, commath,amsfonts}
\usepackage[caption=false]{subfig}

\usepackage{xcolor}

\usepackage[skip=4pt]{caption}

\usepackage{epigraph}
\newcommand{\mrm}{\mathrm}
\newcommand{\RR}{\mathbb{R}}

\def \figpath {figures/}

\begin{document}

\newcommand{\xhdr}[1]{\vspace{0.5mm}\noindent{\textbf{#1.}}\hspace{0.5mm}}

\date{May 2020}
\maketitle


\begin{abstract}
  Many platforms collect crowdsourced information primarily from volunteers. As
  this type of knowledge curation has become widespread, contribution formats
  vary substantially and are driven by diverse processes across differing 
  platforms. Thus, models for one platform are not necessarily applicable to
  others. Here, we study the temporal dynamics of Genius, a platform primarily
  designed for user-contributed annotations of song lyrics. A unique aspect of
  Genius is that the annotations are extremely local --- an annotated lyric may
  just be a few lines of a song --- but also highly related, e.g., by song,
  album, artist, or genre.

  We analyze several dynamical processes associated with lyric annotations and
  their edits, which differ substantially from models for other platforms.  For
  example, expertise on song annotations follows a ``U shape'' where experts are
  both early and late contributors with non-experts contributing intermediately;
  we develop a user utility model that offers one explanation for such behavior. 
  We also find several traits appearing early in a user's lifespan 
  of contributions that distinguish (eventual) experts from non-experts. 
  Combining our findings, we develop a model for early prediction of user expertise.
\end{abstract}


\section{Crowdsourced Lyric Annotation}\label{sec:introduction}

\epigraph{\textit{``Lookin' around and all I see\\ Is a big crowd that's product of me''}}{ --- Kendrick Lamar, \textit{A.D.H.D}}

{\let\thefootnote\relax\footnotetext{See \url{https://github.com/cptq/genius-expertise} for the dataset and code to reproduce experiments.}} 

Online platforms for crowdsourced information such as Wikipedia and Stack
Exchange provide massive amounts of diverse information to people across the
world. While different platforms have varying information structure and goals,
they share a fundamental similarity: the source of the content is a community of
users who contribute their time and expertise to curate knowledge.
Understanding the users that enable the success of these platforms is vital to
ensure the continual expansion of their utility to the general public, but doing
so requires special consideration of the particular structure and form of
information that users contribute.

The activity and expertise distribution amongst users on these crowdsourced information platforms is often heavy-tailed,
and there is substantial effort in understanding the sets of users that make meaningful and/or voluminous contributions~\cite{Pal11,Movshovitz-Attias13},
and how contributions change over time~\cite{Anderson12}.
One example problem is expertise detection~\cite{Zhang07}.
On platforms such as Quora and the TurboTax live community, experts are not clearly defined but longitudinal data can help identify experts~\cite{Pal11,Patil15}.
In contrast, Stack Exchange users have explicit reputation scores accumulated over time, and
there is a focus on early determination of user expertise, where ex ante predictions leverage user behavior and their temporal dynamics~\cite{vanDijk15,Pal12}.

Here, we take expertise detection and characterization as test problems to
study the temporal dynamics of user contributions on \emph{Genius} (\texttt{genius.com}),
a platform primarily for crowdsourced transcription and annotation of song lyrics.
Genius was launched as \emph{Rap Genius} (after a brief stint as
\emph{Rap Exegesis}) in 2009, focusing on the lyrical interpretation of English
rap songs.\footnote{\url{https://genius.com/Genius-about-genius-annotated}}
Since then, Genius has grown to encompass lyrics in many genres and languages,
and \texttt{alexa.com} ranks the web site 408th in terms of global engagement as
of May 2020.\footnote{\url{alexa.com/siteinfo/genius.com}}

\begin{figure}[t]
	\centering
	\includegraphics[scale=0.55]{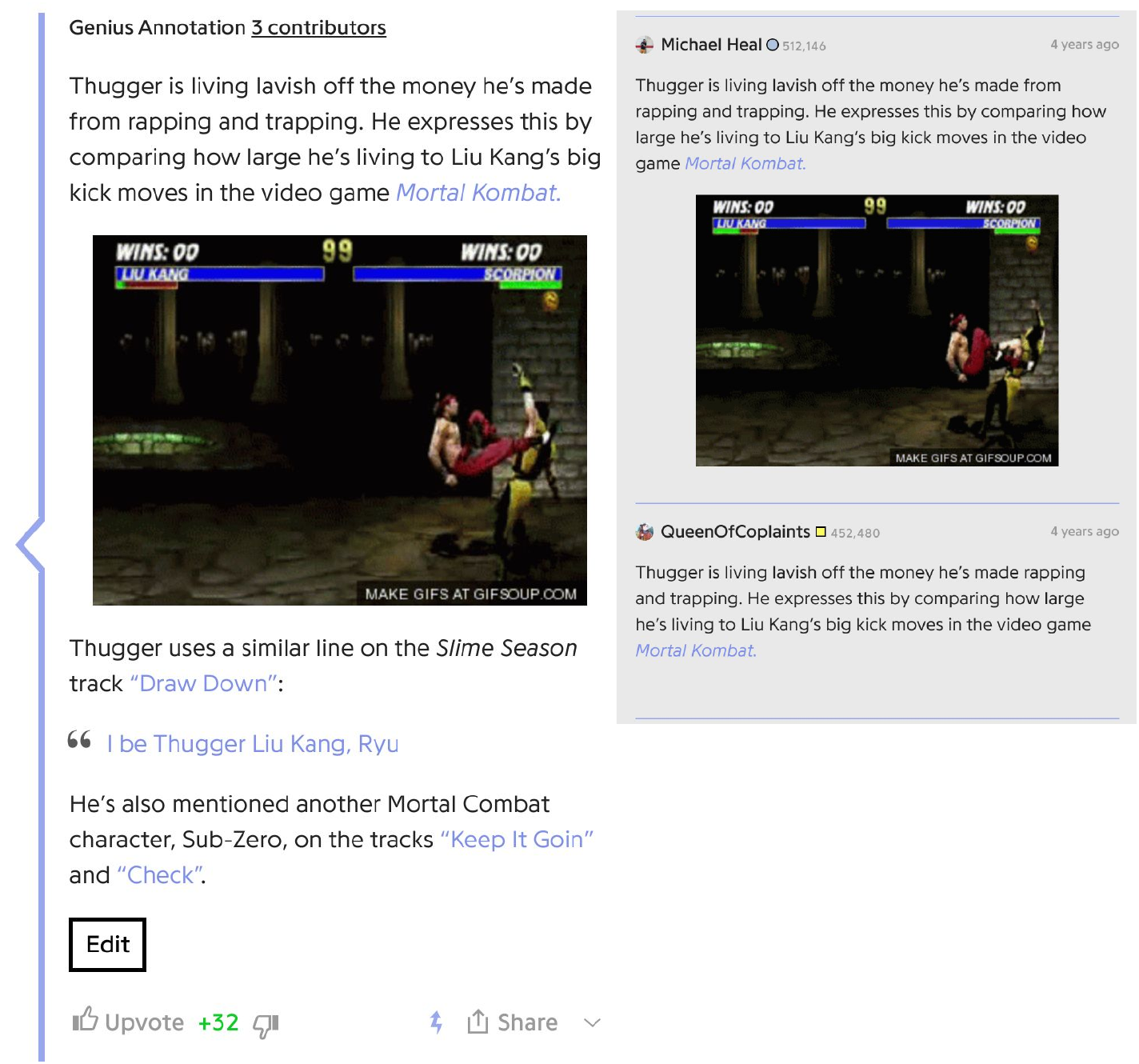}
	\caption{Screenshot of an annotation of the lyrics: ``I'm livin' big, I swear to God I'm Liu Kang kickin'" in Young Thug's song \textit{Digits}
	(\url{https://genius.com/8869644}).
	  (Left) The annotation as of May 2020.
          (Right) Two edits of the annotation. The bottom is the second edit and the top is the third.
	  This annotation has contributions from high-IQ users and contains rich structure such as hyperlinks, a quote, and an image ---
	  all examples
          of what we call \textit{quality tags} (Section~\ref{section:quality}).}
	\label{Screenshot}
\end{figure}

On Genius, lyrics are originally sourced from users who work together to
transcribe songs. After, users annotate the lyrics for interpretation, and the
annotations are edited over time (Fig.~\ref{Screenshot}). The structure of
annotations on Genius differs substantially from other large and popular
crowdsourced platforms such as Stack Exchange, Quora, Yelp, and Wikipedia. For one,
Genius has no explicit question-and-answer format. Instead, the transcription of
newly released songs offers an implicit question generation mechanism, namely,
what is the meaning of these lyrics?  Similar to Wikipedia, annotations are
continually edited to provide a singular authoritative narrative.
However, annotations are extremely localized in the song --- an annotated lyric may be just a
couple of bars in a rap song or the chorus of a folk song. Still, annotations
are related within lyrics of the same song, album, artist, or genre. Finally,
the content is much less formal than Wikipedia, as annotations often contain
slang, jokes, and profanity.

The unique structure of Genius leads to user
dynamics that do not align with existing models for other crowdsourced information sites.
In particular, the partitioning of a song's lyrics
into smaller segments where users choose to annotate or edit content is unique.
While two answers on the same Stack Overflow
question may cover different aspects of the same question, they still result
from the same prompt (the question). Likewise, two Amazon reviews on different properties
of the same product originate from the same prompt (the product). However, two annotations
on the same song on Genius have different prompts --- namely, the two different
lyric segments with which each annotation is associated.
There is also some notion of competition and thus congestion in the choice
of lyric segments to annotate, since there may only be one annotation
per lyric segment, which contrasts with the unbounded number of possible
answers on Stack Overflow or reviews on Amazon products.
Moreover,
while edits on the same annotation come from the same prompt (same lyrics),
the edits are related by the past state of the given annotation.

One part of our analysis is on user \emph{IQ} --- Genius's aggregate measure
of activity and experience of users that is analogous to reputation on Stack Exchange. 
We measure IQ at roughly one point in time and use it as a proxy for (eventual) expertise or experience
when analyzing user behavior retrospectively.
While studies of other crowdsourced information sites apply to user behavior and IQ
on Genius, the structure of annotations also necessitates specialized metrics.
Thus, we define metrics of annotation quality, coverage of lyrics by
annotations, and lyric originality, which help uncover key traits of
user contribution behavior that distinguish experts.

We find several patterns in the temporal dynamics of annotations and subsequent edits on Genius
that have not been observed in prior studies of other crowdsourced information platforms.
One distinguishing set of characteristics are ``U-shaped'' relationships in annotations made on a song over time.
For example, early annotations on a song are made by high-IQ experts (or eventually high-IQ users),
intermediate annotations are made by low-IQ or inexperienced users, 
and the most recent annotations are again made by eventual high-IQ users.
We conceptualize this through an \emph{IQ diamond} ($\Diamond$), in which a song lyric travels from top to bottom, being considered by users that are experts or eventual experts in the narrow top and bottom,
and by the bulk of lower-expertise users in the wider middle.
The quality of annotations and originality of annotated lyrics also follow similar patterns:
the earliest and latest annotations are on average of higher quality and on more original lyrics.

Our IQ diamond model for annotation dynamics contrasts sharply with answer
arrival dynamics on Stack Overflow. For example, \citet{Anderson12} described a
``reputation pyramid'' model of user organization in which questions are first
answered by higher-reputation users and then by lower-reputation users. This
Stack Overflow model does not work as a model for Genius annotations, as it does
not explain the increasing IQ in the later song annotations. Furthermore, the
model does not agree with the editing dynamics; later edits tend to be made by
more experienced users and increase the quality of the annotation.
Similarly, review quality on Amazon and Yelp has a negative relationship
with time \cite{bai2018characterizing},
which disagrees with the ``U-shaped'' relationship of Genius annotation quality
with time and the positive relationship of Genius edit quality with time.

To explain the materialization of the IQ diamond and to further understand
latent factors inducing user annotation patterns, we develop a model of user
utility based on network effects and congestion. In this model, users of
different experience levels gain different utility from creating new annotations
on a song, given the fraction of lyrics that are already annotated.
Fitting simple parametric functions for network effects and congestion matches
empirical differences in user behavior between low-IQ and (eventual) high-IQ users.
This model offers one explanation for behavior, and we discuss alternative ideas
such as loyalty, which is a driver of behavior on Reddit~\cite{hamilton2017loyalty}.

Similar to studies on Stack Overflow~\cite{Movshovitz-Attias13} and
RateBeer~\cite{DNM13,McAuley13}, we also analyze user evolution
stratified by eventual IQ levels. We find inherent traits of eventual experts
visible in their early contributions: even in their first annotations on the
site, eventual experts create higher quality annotations, are more often an
early annotator on a song, are more often an early editor of an annotation, and
annotate lyrics that are more original.
We use these features to design a simple
discriminative model that successfully makes early predictions of eventual
super experts --- users with very high IQ on the site.

\subsection{Additional Related Work}\label{section:Related}

Basic statistics of Genius have been reported with
models for the trustworthiness of annotations~\cite{AlQundus18,AlQundus18_2}.
Additionally, the platform is  used for African American literature courses~\cite{Rambsy18} and for understanding music history~\cite{Dawson18}.
Other music databases --- such as CDDB/Gracenote, MusicBrainz, and Last.fm --- have been analyzed in the context of recommender systems and information retrieval~\cite{swartz2002musicbrainz,celma2010music,koenigstein2011yahoo,schedl2014music}.
More broadly, numerous types of annotation systems have been studied~\cite{Kalboussi15}.

In this paper, we also analyze Genius in the context of other well-studied crowdsourced information sites, such as
Stack Overflow~\cite{ravi2014great,Posnett12,tian2013towards},
Quora~\cite{Wang13,maity2015analysis},
Yahoo Answers~\cite{Adamic08},
Amazon~\cite{bai2018characterizing},
and
Wikipedia~\cite{beschastnikh2008wikipedian,Mesgari15}.
The temporal dynamics of user activity on such sites have been
studied in several contexts~\cite{Anderson12,Jurgens12,Paranjape17,Patil15,almeida2007evolution}.

Expertise plays a central role in the study of crowdsourced information.
For example, employee-labeled ``superusers'' have been studied in the TurboTax live community~\cite{Pal11},
and Quora's manually identified ``Top Writers'' can be identified by temporal behavior~\cite{Patil15}.
On Stack Overflow, early user activity and textual features have
been used for predicting eventual experts~\cite{Movshovitz-Attias13,vanDijk15,Pal12}.


\section{Data Description}\label{DataDescription}

We crawled \texttt{genius.com} from September 2019 to January 2020,
though most of this crawl consists of data that we did not use in
this work. The data spans over 10 years of user activity,
with the earliest activity in October 2009.
We refer to any page that can be annotated as a \emph{song}, even though a small fraction of these pages correspond to other
content such as transcriptions of podcasts or parts of a movie.  Users create
\emph{annotations} on specific lyric segments within a song.
In this work we use `lyrics' to generally refer to words in a song,
and `lyric segment' to refer to the sequence of words that
belong to a single annotation. A lyric segment may consist of multiple lines
of lyrics, or it may even be as small as to only include part of a line.
When there are unannotated lyrics on a song page, a Genius user
can choose a subset of unannotated lyrics to annotate (thus creating a lyric segment).
After a user makes an initial \emph{annotation}, users can create an
\emph{edit} for an annotation, and the most recent version of the annotation (with edits) is
displayed on the web page.
We typically use ``edits'' to refer to the changes in annotations
and reserve the word ``annotation'' for the initial annotation.

\begin{figure}[t]
  \centering
    \includegraphics[width=0.48\columnwidth]{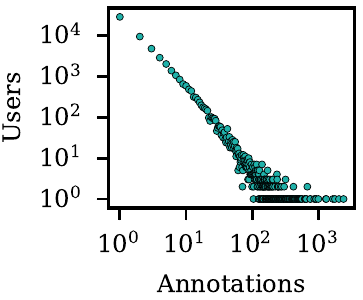} \hfill
    \includegraphics[width=0.48\columnwidth]{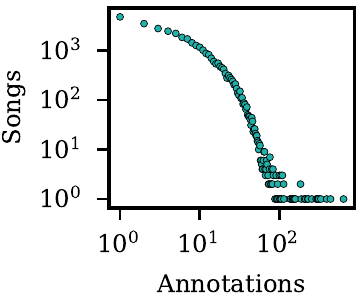}
  \caption{(Left) Distribution of annotation counts of users.
    (Right) Distribution of annotation counts on songs.}
  \label{AnnEditDistr}
\end{figure}

We collected lyrics, annotations, and annotation edit history from 223,257 songs.
Of these, 33,543 songs have at least one annotation. 
In total, we collected 322,613 annotations and 869,763 edits made by 65,378 users.
For each annotated lyric segment, we have the complete timestamped history of edits and content.
The annotations and edits consist of text and HTML.
We find that the distribution of the number of annotations made by a user
and the number of annotations on a song are heavy tailed (Fig.~\ref{AnnEditDistr}).

For each user, we have every annotation and edit that they have made
on the collected songs.
We also recorded the \emph{IQ} of each user \emph{at the time of our crawl}.
IQ is an aggregate measure
that accounts for various contributions on Genius, such
as writing annotations and transcribing songs.
As we only have snapshots of IQ and not the evolution of IQ over time,
our analysis is based on the IQ that users have accumulated at the time
of data collection, and we use this to study user behavior retroactively. 
Thus, when we refer to a high-IQ user in later sections,
we mean that this user \emph{eventually} accrues a large IQ score by the time that
we have collected this data.
In other words, when analyzing user behavior retroactively, such users could have had high IQ
at the time of an action, or they eventually earned high IQ by the time we collected the data.
Users with low IQ have always had low IQ.
When analyzing annotation dynamics and expertise in later sections,
we often have to restrict to
users with sufficient data (i.e., users with sufficiently many annotations) and
these users are often old enough on the site so that their IQ has
stabilized and their experience can be approximately measured by IQ
(see user age statistics in Section~\ref{sec:evolution}).
However, this type of analysis limits some of our user behavior findings,
as we do \emph{not} have IQ measurements at the time of the user activities,
and this should be kept in mind when interpreting the results.

One way users accumulate IQ is when their annotation gets upvotes from other users.\footnote{https://genius.com/8839950}
Moreover, users can be given certain roles in the Genius community, 
such as Editor, Moderator, and Mediator; these roles come with certain permissions
and signal certain accomplishments of the user.\footnote{https://genius.com/Genius-what-is-a-moderator-annotated} 
The amount of IQ earned for an upvote is higher if that upvote comes from a user with one of these roles.
An annotation made by a user can result in a net loss of IQ if it is sufficiently downvoted,
but we do not know how much net IQ is gained per annotation or
the users that vote on an annotation. We do not separate
users by role, nor we do not account for pressure against annotation
due to possible loss of IQ.

In addition to annotations and edits, we also collected other Genius user
actions such as suggestions, questions, answers, comments, and
transcriptions. We say that a user has \emph{contributed} to a song if they have
recorded some interaction with the song. 
However, our analysis focuses on annotations and edits.
Although we do not have the exact percentage of user IQ
that is obtained from annotations or edits, we believe that it accounts for a large portion of IQ
for many users.
User IQ and annotation counts are highly correlated ($r = 0.633$), and user IQ follows a
similar distribution to that of user annotation counts.


\section{Metrics for Annotations}

The annotations that users create and edit on Genius are a unique form of crowd
contribution. In order to study user behavior and dynamics, we first define
metrics for annotations related to quality, coverage, and originality.

\subsection{Annotation Quality}\label{section:quality}
To better understand the generation of content on Genius, we would like to quantify annotation quality.
We simply take the number of HTML tags that indicate rich
content creation as one proxy for quality. Specifically, we consider the following \emph{quality tags}:
\texttt{<a>}, \texttt{<img>}, \texttt{<iframe>}, \texttt{<blockquote>},	\texttt{<twitter-widget>}, \texttt{<ul>}, \texttt{<ol>}, and \texttt{<embedly-embed>}.
The annotation in Fig.~\ref{Screenshot} has three unique quality tags: \texttt{<blockquote>}, \texttt{<img>}, and \texttt{<a>}. 
Through manual inspection, we find that the presence of these tags tends to indicate
helpful or interesting information in annotations,
making the number of such tags a useful and simple
proxy for annotation quality on Genius.
In Section~\ref{sec:evolution}, we show that higher quality tag counts in
annotations distinguishes high-IQ users, even on their earliest annotations on the site.
Moreover, in Section~\ref{sec:prediction}, quality tag counts are shown to be an informative feature for classifying super experts.

Many of these quality tags are associated with quality of
user-generated content in other sites.  For instance, featured articles on
Wikipedia have more links and images than other articles~\cite{Stvilia05}, the
probability of answer acceptance on Stack Overflow positively correlates with
the number of URLs in the answer~\cite{Calefato15}, and the probability of
retweeting on Twitter positively correlates with the presence of URLs in the
tweet~\cite{Suh10}.  

We also measure annotation quality by \emph{length}, i.e., the number of characters in the content.
This has been an effective metric for content quality on Wikipedia, Yahoo Answers, and Stack Exchange~\cite{Blumenstock08,Stvilia05,Adamic08,Gkotsis14}.
Later, we detail nuances in using annotation length as a proxy for annotation quality, as 
we find pressure to annotate early may cause the first annotations on a song to be shorter.

\subsection{Annotation Coverage}

\begin{figure}[t]
  \centering
  \includegraphics[width=0.48\columnwidth]{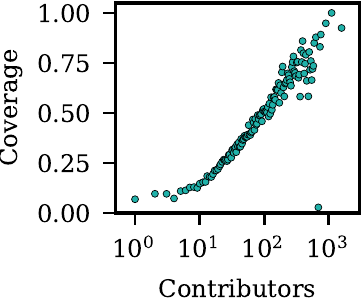} \hfill
  \includegraphics[width=0.48\columnwidth]{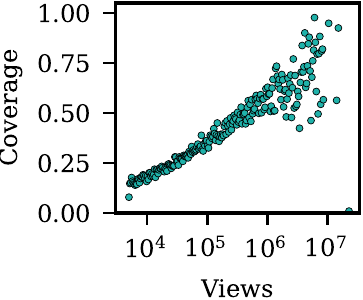}
  \caption{Mean annotation coverage as a function of number of contributing
    users on a song (left) and number of page views on a song (right), computed
    over songs with at least one annotation. Increased popularity with site
    contributors and visitors correlate with higher coverage.
    (Note: Genius only lists view counts for songs with at least 5,000 views).}
  \label{CoverageSongStat}
\end{figure}

The extent to which crowdsourced contributions satisfy the needs of information
seekers is vital to a platform's success. To this end, we consider
\emph{coverage} --- the amount of the information sought by visitors that is
actually present on the site. Various notions of coverage have been used for
analyzing Stack Overflow discussions~\cite{Parnin12}, as well as
accuracy and topical concerns of
Wikipedia~\cite{Mesgari15,Giles05,Samoilenko14,Brown11}

As the fundamental function of Genius is to provide annotations on documents, we
study coverage of lyrics by annotations. While some lyrics may be fillers or
seemingly lack meaning, there is often still potential for interesting
annotations. For example, annotations on such lyrics may provide references to
other similar lyrics, references to related external social media content, or
historical context.

Starting from all the lyrics of a song, we compute its \emph{annotation coverage} as follows.
First, we remove lyrical headers (e.g., ``[Verse 1: Kanye West]'' or ``[Refrain]'').
This leaves $L$ total text characters available for coverage.
Next, let $A$ be the total number of text characters in all
lyric segments that have been annotated.
Then the \emph{annotation coverage} for the song is $A / L$.
We note that this accounts for repeated sections of the song, since if
for instance there are 3 repetitions of a chorus, then any annotation
on a lyric segment in the chorus will usually also be considered
an annotation on the 2 repetitions of this lyric segment in the other parts of the song.
We find a positive relationship between annotation coverage and both the number of users
contributing to the song as well as the number of views of a song (Fig.~\ref{CoverageSongStat}).

\subsection{Lyric Originality}

\begin{figure}[t]
  \centering
  \includegraphics[width=0.7\columnwidth]{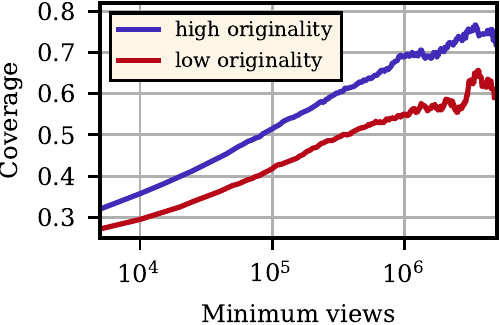}
  \caption{Mean annotation coverage over songs as a function of number of page views,
    stratified by songs in the upper third of originality (blue), and songs in the lower third of originality (red).
    Songs with more original lyrics tend to have higher annotation coverage.
  }
  \label{AnnCoverageComplexity}
\end{figure}

We find that annotation coverage also depends on the originality of the lyrics of a song.
To measure originality, we first compute inverse document frequencies (idfs), where documents are songs, i.e.,
$\textnormal{idf}(w) = \log (N / \textnormal{df}(w))$, where $N$ is the total number of songs and $\textnormal{df}(w)$ is the number of songs containing word $w$.
These words $w$ are pre-processed by the same procedure as done for annotation coverage.
Following ideas from \citet{Ellis15}, we define the \emph{originality} of lyrics $\ell$ appearing in a song as the following $L$-estimator:
\begin{equation}\label{eq:originality}
\mrm{originality}(\ell) = (p_{60} + p_{75} + p_{90}) / 3,
\end{equation}
where $p_{x}$ denotes the $x$th percentile of the idf values of unique words in
$\ell$ (with linear interpolation between percentiles if needed).
We use large percentiles as such words are of interest to site
visitors and annotators, and many words in song lyrics may be fillers for
aesthetic reasons (e.g., leading to a rhyme). Only computing percentiles over
unique words prevents long, repetitive songs from achieving high originality
scores. We find that more original songs tend to have higher annotation
coverage (Fig.~\ref{AnnCoverageComplexity}).

To illustrate, one lyric with low originality score is ``I can never make him love me / Never make him love me'' (Frank Ocean, \textit{Bad Religion}), which has a score of 0.927. Note that each word of the second line is contained in the first line, so only the words of the first line are used to compute originality. Each word appears in many other songs, so the lyric has low originality. An example of a lyric with high originality is ``a solo is now a poet / Hypnosis overdose on potions, adjustin' to the motions / And gettin' out all my emotions'' (A\$AP Rocky, \textit{Everyday}), which has a score of 5.477. The words of highest idf are `adjustin' at 8.06, `hypnosis' at 7.91, and `potions' at 7.50. The 60th, 75th, and 90th percentiles of idf are 3.47, 5.38, and 7.58.


\section{Dynamics of Annotations and Edits}\label{sec:annotation_dynamics}

We now investigate relationships involving the temporal order of annotations
and edits. We define the \emph{time rank} $R$ of an
annotation on a song as the numerical position in which it was created; for
example, the third earliest annotation on a song has a time rank of 3. Similarly, 
we define the \textit{proportional time rank} $q$ of an annotation for a song with at least two
annotations by $q = (R-1)/(n-1)$, where $n$ is the number of
annotations on the song. This measurement allows for comparison of annotations
with similar relative positions in a song's lifespan across songs with different
numbers of annotations.
The number of annotations on fully annotated songs (where every part of the lyrics is annotated) 
varies substantially, and some songs are not fully annotated yet have reached a type of equilibrium state
of annotation coverage in which all lyric segments that have meaningful
annotations have been annotated.
One of these two cases seems to hold for most songs in our dataset,
although we can never be sure if a song that is not fully annotated is truly ``in equilibrium'',
as annotations may arrive after the point in time when we collect data.
Comparing annotations by proportional time rank is generally robust to 
all of this variation in the data.
We use analogous definitions for edits on an annotation,
and we define the edit at time rank $0$ to be the (initial) annotation itself.
Edits have similar variation, as there is no way to tell whether an annotation
has reached its final edited state.

Below, we analyze how temporal orders relate to both contributing users and to the content
itself. One of our findings is that annotator experience and measures of annotation quality
exhibit a ``U-shaped'' pattern with respect to the proportional time rank of
\emph{annotations}. In other words, more users that already have or will
eventually have high experience (i.e.,
have high IQ at the time we collected the data) and higher-quality
content appear both early and late in time, with less experienced users (i.e., those with low IQ) 
and lower quality content in the intermediary. This is different from
the negative relationship over time of user reputation for Stack Overflow answerers~\cite{Anderson12}, 
and the negative relationship over time of review quality on Amazon and Yelp~\cite{bai2018characterizing}.
We develop an ``IQ diamond'' model of user behavior based on ideas of economic utility
that offers one explanation for this behavior, while also discussing alternative explanations.
In contrast, for \emph{edits}, eventual experience and quality grow over time.

\subsection{Dynamics of Annotations}\label{sec:annotationdynamics}

\begin{figure}[t]
  \centering
  \includegraphics[width=0.45\columnwidth]{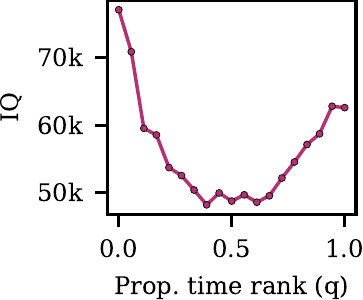} \hfill
  \includegraphics[width=0.45\columnwidth]{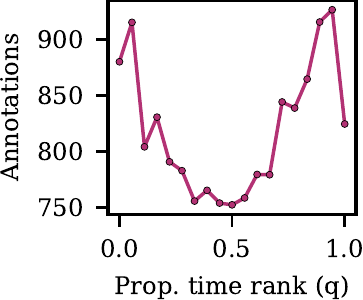} \\[3mm]
  \includegraphics[width=0.45\columnwidth]{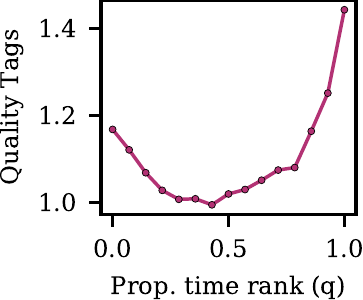} \hfill
  \includegraphics[width=0.45\columnwidth]{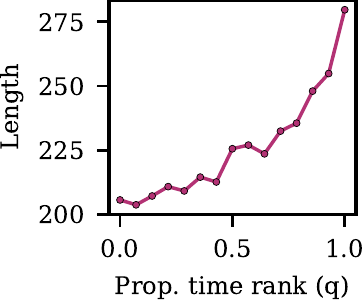}
  \caption{Various user and content statistics as a function of the proportional
    time rank of an annotation. Mean IQ (top left), total number of annotations
    made by the annotating user (top right), and number of quality tags (bottom
    left) follow a ``U shape'' with respect to proportional time rank.}
  \label{AnnotTimeStat}
\end{figure}

First, we analyze temporal dynamics of annotation creation (we do not consider edits
until Section~\ref{sec:edit_dynamics}).
Measuring the (eventual) IQ of a user making an annotation as a function of the proportional time rank
(Fig.~\ref{AnnotTimeStat}, top left), we find the aforementioned ``U shape''.
Early annotations on a song are made by high-IQ users.
After, the mean IQ
decreases monotonically with proportional time rank until around half the total
annotations have been made. After these middle annotations, the mean IQ increases
monotonically. 
(We remind the reader that this is IQ at the time of data collection and not necessarily
at the time of edits. Thus, high-IQ users either had high IQ at the time or eventually accrued high IQ,
whereas low-IQ users always had low IQ.)
This differs from other platforms such as Stack Overflow,
where there is a monotonic
decrease in user reputation over time~\cite{Anderson12}. We also find that the
number of total annotations made by a user follows the same trend
(Fig.~\ref{AnnotTimeStat}, top right). Thus, the ``U shape'' is present if we measure
experience of users by an aggregate measure such as IQ or simply the total number of
annotations the user has made.

We also consider annotation quality as a function of proportional time rank
(Fig.~\ref{AnnotTimeStat}, bottom row). Here, the number of quality tags follows the
same ``U shape''. Thus, not only are the earlier and later annotations on songs
made by more able users, they are also of higher quality under
this metric. However, our other quality metric --- annotation length ---
largely increases with proportional time rank. A possible contributing factor
is that early annotators may feel time pressure to annotate lyrics before
others, incentivizing shorter annotations that are faster to create. This trend
indicates that annotation length measures other factors beyond annotation
quality. Such time pressure may also explain the somewhat lower number of
quality tags for the earliest annotations compared to the latest annotations.
We note that the earliest reviews on Amazon and Yelp are often more helpful and longer,
with later reviews being of lower helpfulness and length~\cite{bai2018characterizing}. 
The positive trend of annotation length over time rank
on Genius has the opposite trend, and the ``U-shaped'' trend for quality tags is
of a different nature.
\begin{figure}[t]
	\centering
	\includegraphics[scale=1]{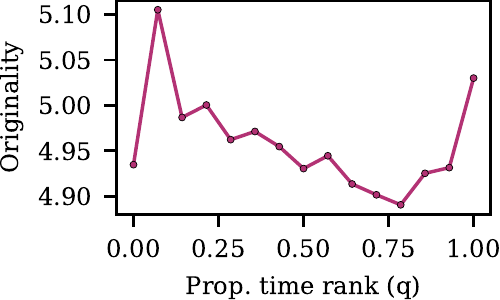}
	\caption{Lyric segment originality as a function of annotation proportional time rank.
          Apart from the first annotation, originality follows a ``U shape''.}
	\label{AnnotTimeComplexity}
\end{figure}
\vspace{-10pt}

Finally, we consider how annotated lyric segment originality relates to temporal annotation
ordering. Again, for the most part, we see the familiar ``U shape''
(Fig.~\ref{AnnotTimeComplexity}), indicating that the complex lyrics are
annotated both at the beginning and end. However, the first annotations
tend to be on lyrics with lower originality. This may be due to users'
reluctance to annotate before there are any other annotations, leading them to
annotate simpler lyrics to get the ball rolling.

\subsection{The IQ diamond ($\Diamond$)}\label{sec:diamond}
At a high level, annotation dynamics on Genius appear to be markedly different
from dynamics on Stack Overflow, Amazon, and Yelp.
\citet{Anderson12} observed that user reputation decreases
with later answers to a given question on Stack Overflow. From this, they
develop a ``reputation pyramid'' model of user behavior, where new questions are
first considered by high-reputation or experienced users before being considered
by users with less experience. On Genius, we see the same initial descent but
then a curious ascent producing the ``U shape.'' On average, early annotators are
experienced users (or users that will eventually become experienced, i.e., high IQ by the time we collected the data),
who quickly make annotations of high quality and on more novel lyrics. 
After, users with low experience levels make annotations of typically lower
quality on lyrics that are less novel. Finally, the late annotations are again
made by experienced users on the remaining lyrics, which tend to be more
original. This behavior suggest an \emph{IQ diamond} ($\Diamond$) model for
Genius, in which song lyrics are first processed by high-IQ users that form a
narrow point of the diamond, then the song opens to a broader set of users to
form the wide middle of the diamond, and finally narrows to the high-IQ users again.

To explain the IQ diamond pattern, we develop a model of utility for user
annotation. The utility of annotating a given song
at any time depends on the
proportion of the song that is already annotated at that time. More annotation
coverage impacts utility both positively and negatively --- one can gain more IQ
by annotating songs with more activity, but higher coverage limits the choice of
lyrics that a user may annotate. Thus, the users' utility functions may be
modeled similarly to the utility of services with both (positive) network effects and
(negative) congestion effects~\cite{Johari09}. Related models have been employed
for users on crowdsourcing systems~\cite{Chen19}.

A Genius-specific feature of our model is that it considers user
contribution to a set of prompts (lyrics) that are all related
as lyrics of the same song. As mentioned in the introduction,
this structure distinguishes Genius from other sites with different contribution
dynamics. Any two separate annotations on the same song on Genius originate from different
prompts (the two separate lyric segments) and can cause congestion (since others
cannot annotate these lyric segments), whereas two reviews on an Amazon product
answer the same prompt and do not prohibit others from adding related reviews~\cite{gilbert2010understanding}.
Our model provides one possible (but interesting) explanation of the unique
user behavior on Genius.
We discuss possible alternative mechanisms and limitations of our model at the end of this section.

More specifically, first fix a song, and consider a population of $N$ users labeled $1$ to $N$.
Let $\rho \in \RR^N$ be the vector for which the $k$th entry $\rho_k$ is the annotation coverage of the song by user $k$'s annotations, so $\rho_k \geq 0$ and $\sum_{j=1}^N \rho_j \in [0,1]$. Here we will refer to an annotation as an infinitesimal increase in coverage. We model the expected utility that user $k$ would derive from adding an annotation by
\begin{equation}\label{eq:utility}
\textstyle u_k(\rho) = b_k + f_k\left(\sum_{j \neq k} \rho_j \right) - g_k\left(\sum_{j=1}^N \rho_j \right),
\end{equation}
where $b_k$, $f_k$, and $g_k$ are such that
\begin{itemize}
	\item $b_k \geq 0$ is the expected a priori personal utility that user $k$ derives from annotating a random lyric segment 
	(we are assuming here that users never have negative utility from annotating).
	
	\item $f_k(x) \geq 0$ is a nondecreasing function measuring the expected positive network effect when $x$ proportion of the lyrics are covered by other users.
          The positive network effect arises because users tend to gain more IQ and have their contribution viewed by more people on songs that are more popular.
          Empirically, coverage is positively correlated with the number of song page views (Fig.~\ref{CoverageSongStat}). 
	\item $g_k(x) \geq 0$ is a nondecreasing function that measures the expected congestion effects when $x$ proportion of the lyrics are covered; lyrics that are already annotated cannot be annotated by user $k$.
\end{itemize}

For simplicity, we only consider two types of users: high-IQ ($h$) and low-IQ
($l$). For subsequent measurements, we consider high- and low-IQ users as those in the top
third or bottom third in IQ (as measured at the time data was collected) over all users with at least 10 annotations.

Suppose that some user $k$ has not yet annotated a song.
Then $\sum_{j \neq k} \rho_j = \sum_j \rho_j$ is the annotation
coverage and is just a proxy for the proportional time rank in our
infinitesimal setting. Assuming the likelihood that user $k$ makes an annotation
is proportional to their utility $u_k(\rho)$, we would see that a
user would make annotations at proportional time ranks corresponding to
points at which their utility is high.

\begin{figure}[t]
  \centering
  \includegraphics[width=\columnwidth]{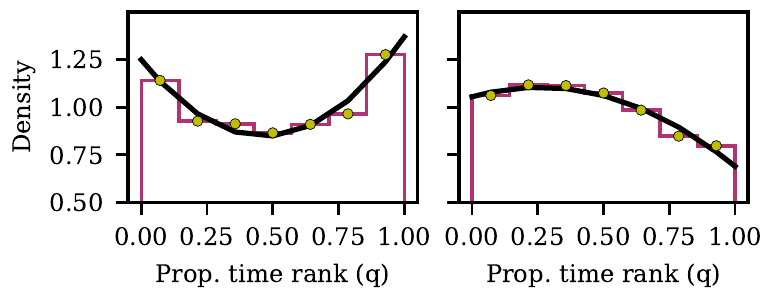}
  \caption{Distribution of proportional annotation time ranks for high-IQ (left) and low-IQ (right) users.
    The red bars are histogram bins where the height is the probability density.
    The black curves are our utility models that are fitted to the histogram bin midpoints (shown as yellow points).
  }
  \label{PropTimeRankDistr}
  \vspace{-20pt}
\end{figure}

We measure this empirically by simply considering the distribution of
proportional time ranks of annotations made by a high- or low-IQ user
(Fig.~\ref{PropTimeRankDistr}). Next, we fit the model in Eq.~\ref{eq:utility}
to these approximate utility curves. To this end, we make some assumptions on
$f_k$ and $g_k$. When there are no annotations, there are no network or
congestion effects, so $f_k(0) = g_k(0) = 0$. Both of these functions are
nonnegative and nondecreasing. 
We will assume that $f_k$ and $g_k$ are both concave, which
could be due to due to diminishing returns~\cite{Chen19}.
One extremely simple class of concave functions are quadratics, so we
fit $f_k$ and $g_k$ as quadratic functions. Under our assumptions, the model parameters satisfy
\begin{align}
	f_k(x) & = -a_1^{(k)} x^2 + a_2^{(k)} x, & a_1^{(k)} \geq 0, a_2^{(k)} \geq 2a_1^{(k)} \label{constr1}\\
	g_k(x) & = -c_1^{(k)} x^2 + c_2^{(k)} x, & c_1^{(k)} \geq 0, c_2^{(k)} \geq 2c_1^{(k)}. \label{constr2}
\end{align}
To determine these coefficients, we take the histogram midpoints $x_1, \ldots, x_t$ and histogram heights $y_1, \ldots, y_t$ from the proportional time rank distribution (Fig. \ref{PropTimeRankDistr}) and solve the linear least squares problem
\begin{equation} 
\textstyle \min_{b_k, a_j^{(k)}, c_j^{(k)}} \sum_{j=1}^t (b_k + f_k(x_j) - g_k(x_j) - y_j)^2,
\end{equation}
subject to the constraints (\ref{constr1}), (\ref{constr2}), and $b_k \geq 0$.
We solve this problem for $k \in \{h, l\}$ for the two sets of users.
Table~\ref{utilityCoeffs} shows the fitted coefficients and Fig.~\ref{PropTimeRankDistr} shows the resulting curves, which match the empirical distribution.

\begin{table}[t]
  \caption{Fitted parameters of utility functions.}
  \begin{center}
    \begin{tabular}{r c c c c c}
      \toprule
      & $b$ & $a_1$ & $a_2$ & $c_1$ & $c_2$ \\
      \midrule
      high-IQ (h) & 1.25 & 0.003 & 2.02 & 1.84  & 3.74 \\
      low-IQ (l)  & 1.06 & 0.79 & 1.83 & 0.04 & 1.44 \\
      \bottomrule
    \end{tabular}
  \end{center}
  \label{utilityCoeffs}
\end{table}

The coefficients in Table~\ref{utilityCoeffs} match the IQ diamond model and give evidence
for some more specific properties of user contribution behavior.
First, $b_h > b_l$, which is sensible as the a priori utility that high-IQ users receive for annotating should be higher.
Indeed, high-IQ users may derive extra benefits due to their status in Genius's social network or increased attention on other social media accounts linked to their Genius profile.
Also, $a_2^{(h)} < c_2^{(h)}$ while $a_2^{(l)} > c_2^{(l)}$.
Since $f_k'(0) = a_2^{(k)}$ and $g_k'(0) = c_2^{(k)}$, these inequalities imply that when a song only has a few annotations,
high-IQ users are more influenced by the loss in utility from the few already existing annotations (congestion) while low-IQ users are more positively influenced by the 
presence of those few annotations (network effects).

Since $0 \approx a_1^{(h)} \ll a_1^{(l)}$, network effects in the fitted model
approximately scale linearly for high-IQ users while they are most significant when there are few annotations for low-IQ users.
This could arise if network effects for high-IQ users are due mostly to IQ gains from additional views and activity on a song. Recall that users earn IQ per upvote on their annotation, so the linearity of network effects may be due to the approximately linear positive relationship between song views and upvotes ($r=0.515$).
On the other hand, network effects for low-IQ users might come from social factors.
The larger marginal network effects felt for early annotations may be due to desire of low-IQ users to achieve a baseline social validation that the song is worth annotating.

For congestion effects, $0 \approx c_1^{(l)} \ll c_1^{(h)}$. Thus, in our fitted model,
congestion approximately scales linearly for low-IQ users while it is most significant when
there are few annotations for high-IQ users. This could be caused by high-IQ users
having more selectivity about the lyrics that they choose to annotate. This would agree
with experienced reviewers on Amazon products that are opinionated in 
their reviews~\cite{gilbert2010understanding}.

In our model, one could decompose the congestion
effect by $g_k(x) = g_{k,s}(x) + g_{k,g}(x)$, where $g_{k,s}(x)$ is the expected
utility lost due to lyrics that user $k$ is qualified to annotate
or is especially interested in annotating having already
been annotated, and $g_{k,g}(x)$ is the utility lost due to other lyrics already
being annotated. If $g_{k,g}(x)$ is proportional to the amount of general
knowledge that user $k$ has to annotate lyrics, we can assume it is linear.
Then by concavity of $g_k(x)$, the function $g_{k,s}(x)$ is also concave.
With this
decomposition, a sufficient condition for the observed inequalities 
$c_1^{(l)} < c_1^{(h)}$ and $c_2^{(l)} < c_2^{(h)}$ is that high-IQ users have both more
specific knowledge and general knowledge about lyrics than low-IQ users. The low
value of $c_2^{(l)}$ would then suggest that low-IQ users have little specific knowledge and
are not particularly selective in which lyrics to annotate.

\xhdr{Additional considerations and alternative explanations}
There are several additional considerations that may impact temporal dynamics
for which our IQ diamond model does not take into account.
One example could be a form of loyalty in which users interact with songs of their favorite artists,
as loyalty is known to be a driver of user activity on other platforms~\cite{hamilton2017loyalty}.
Artist pages on Genius have leaderboards that list users who have earned much IQ from annotating
the lyrics of a particular artist. Users also have an artist leaderboard on their personal page,
where they can display their positions on the artist leaderboards.
Thus, users may be incentivized to focus their annotations on specific artists to acquire these accolades.
Reddit users that are loyal to a particular subreddit contribute to many
less-popular posts~\cite{hamilton2017loyalty}, so the analog for Genius might be that
users loyal to an artist may annotate many less-popular lyrics. Also,
experienced reviewers on Amazon strongly care about their own brand or identity
as a reviewer~\cite{gilbert2010understanding}, so high-IQ users on Genius may be
more loyal when they have positions on leaderboards.
The effect of loyalty on annotation dynamics may in fact result in
differences between high-IQ and low-IQ users that are not accounted for in our model.

Also, our IQ diamond model only distinguishes users based on IQ and models each
annotation as having the same impact on network effects and congestion,
regardless of the user that made the annotation. Power dynamics between users
may impact temporal dynamics of annotations and edits, as suggested by the
positive relationships between edit time rank and user IQ
(Fig.~\ref{EditTimeStat}, top left). Moreover, some users have various
``roles'' in the Genius community 
(see Section~\ref{DataDescription}), which
may add to imbalanced power dynamics.

The IQ diamond model is also in some ways limited by the data that we have.
Since the model fits the network effects function to data by approximating
the coverage as $\sum_{j \neq k} \rho_j = \sum_j \rho_j$,
we do not consider the behavior of a single user returning to edit an annotation
or adding more annotations to a song.
Also, we do not have the exact state of a user at the time of making each annotation, 
nor do we have sufficient data to determine the exact amount of IQ gained per annotation. As
a user can lose IQ on a sufficiently downvoted annotation, potential IQ loss likely
influences users in ways that are not explicitly captured by our model.
More specifically, the assumption that the parameter $b_k$ is always nonnegative might be unrealistic.

\subsection{Dynamics of Edits on an Annotation}\label{sec:edit_dynamics}

The temporal dynamics of edits are quite different from that of annotations.
Around 95\% of annotations have 9 edits or fewer, so we directly study time ranks instead of proportional time ranks.
We find that users often edit their own annotations, and users often make several consecutive edits in a row.
Removing these edits gives qualitatively similar results, so we present results over all edits.
Recall that in this section we consider the action at time rank $0$ to be the initial annotation (the activity studied in the previous sections), serving as a reference point, and the actions at time ranks strictly greater than $0$ are edits.

\begin{figure}[t]
  \centering
  \includegraphics[width=0.45\columnwidth]{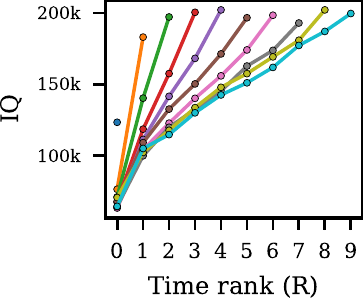} \hfill
  \includegraphics[width=0.45\columnwidth]{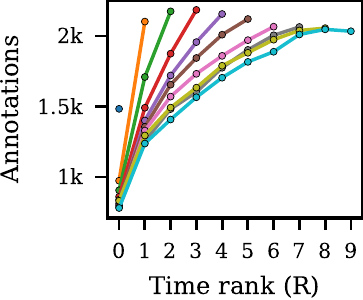} \\[2mm]
  \includegraphics[width=0.45\columnwidth]{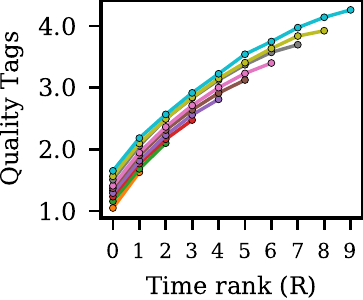} \hfill
  \includegraphics[width=0.45\columnwidth]{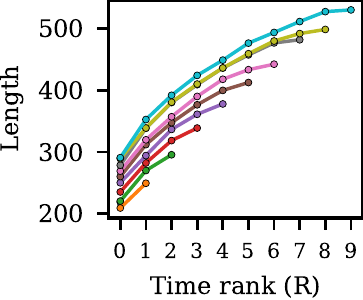}

  \caption{User and content statistics as a function of edit time rank. Each line
    corresponds to a different total number of edits for an annotation. For instance,
	the light blue line ends at time rank 9, so it is the strata with all annotations
	that have 9 total edits. The third dot on the light blue line in the top left figure
	displays the mean user IQ of users making the second edit on annotations with 9 total edits.
	For annotator experience, mean
    user IQ (top left) and number of annotations made by a user (top right) both
    increase with time rank. For quality, the mean number of quality tags
    (bottom left) and mean length (bottom right) increase with time rank.}
	\label{EditTimeStat}
\end{figure}

Figure~\ref{EditTimeStat} shows various properties of edits and the users
making the edits against time rank. Clear relationships emerge when considering
the arrival of edits sequentially over the life of an annotation. Thus, we
separate each plot into 10 strata, each one containing annotations that have $k$
total edits, where $0 \leq k \leq 9$.

We first find that the (eventual) experience of a user --- as measured by both mean IQ and
mean total number of annotations at the time of data collection --- increases as edit time rank increases,
regardless of the number of edits on an annotation (Fig.~\ref{EditTimeStat}, top
row). This positive correlation could be explained by hesitance of users with
less experience to make edits on the content of a user with more experience. The
plots in the top row of Fig.~\ref{EditTimeStat} have the opposite trend of similar
plots for user reputation on Stack Overflow~\cite{Anderson12}. Annotations also
have higher quality with more edits, as measured by both mean number of quality
tags and length (Fig.~\ref{EditTimeStat}, bottom row).
Such a relationship is reasonable as edits are meant to augment content. The plots in the bottom row
of Fig.~\ref{EditTimeStat} show the opposite trend of similar plots of
review length and helpfulness on Amazon and Yelp~\cite{bai2018characterizing}.

There is nuance when comparing annotations with
different numbers of total edits at a fixed time rank (points at a fixed x-axis
value in Fig.~\ref{EditTimeStat}). For a fixed edit time rank, the user making
the edit tends to have higher IQ if there are fewer total edits of the
annotation. However, the edits on annotations with fewer total edits have fewer
quality tags and are shorter. We would expect this behavior in cases when the
initial annotation has more quality tags and longer length, and the complexity
may require more edits for the annotation to reach a final state.


\section{Evolution of User Behavior}\label{sec:evolution}

Having considered temporal dynamics of annotations and edits with respect to
arrivals on a single song or annotation, we now analyze how user behavior
changes over a user's lifespan on Genius. By studying behavior
for users of differing eventual IQ levels, we can better understand how the early and current
behavior of an expert can be distinguished from that of other users. We use
these ideas in Section~\ref{sec:prediction} for early prediction
of super experts. Similar
analysis of user behavior evolution on Stack Overflow has proven useful for
identifying experts based on early behavior~\cite{Movshovitz-Attias13}.

\begin{figure}[t]
  \centering
  \includegraphics[width=0.45\columnwidth]{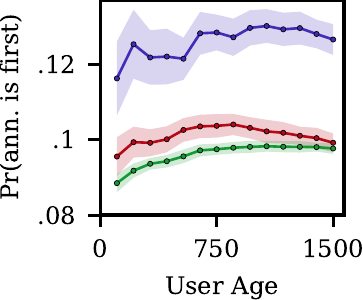} \hfill
  \includegraphics[width=0.45\columnwidth]{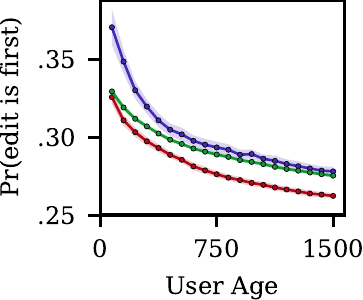} \\[2mm]
  \includegraphics[width=0.45\columnwidth]{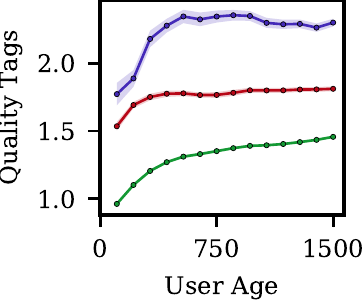} \hfill
  \includegraphics[width=0.45\columnwidth]{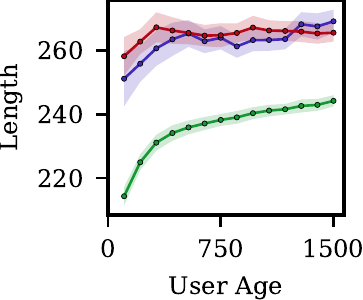} \\[2mm]
  \includegraphics[width=0.45\columnwidth]{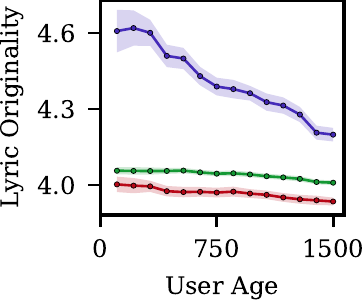} \hfill
  \includegraphics[width=0.45\columnwidth]{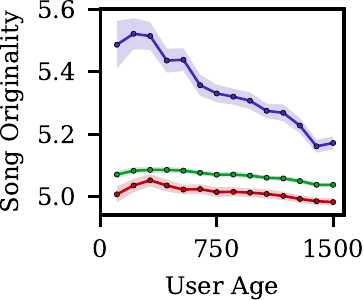} \\
  \caption{User behavior over days since first annotation or edit,
    stratified by high-IQ users (at least 100,000 IQ, blue),
    mid-IQ users (between 10,000 and 50,000 IQ, red),
    and all users (green).
	Shown are cumulative averages estimated by 100 bootstrapped samples at each point,
	with $\pm$ two standard deviations shaded in.
	For instance, the point at (750, 0.128) on the blue line in the top left figure
	indicates that over all annotations made by high-IQ users within 750 days of their
	first annotation on the site, 12.8\% are the first annotation on a song.
    (Top row) 
	For high-IQ users, the proportion of annotations that are the first annotation on a song
	and proportion of edits that are the first edit on an annotation are high.
    %
    (Middle row) High-IQ users make annotations with higher quality tag count, and mid- or high-IQ users
	 make annotations with higher length.
	(Bottom row) High-IQ users annotate lyric segments and songs with higher originality scores.
  }
  \label{Lifespans}
\vspace{-10pt}
\end{figure}

To analyze how user behavior changes over time, we measure cumulative averages
of properties of the annotations and edits that users make over their
lifespans. We also stratify users into three levels based on the IQ that they
have accumulated at the time of our data collection:
users with at least 100,000 IQ
(high-IQ users), users with between 10,000 and 50,000 IQ (mid-IQ), and
a group with all users. In the case of annotations,
we include users with at least 10 annotations and
for each user consider activity over the first 1,500 days after their first
annotation. For edits, we consider users with at least 10 edits
and study the 1,500 days after their first edit.
There are 5,162 total users with at least 10 annotations, of which 176
are in the high-IQ strata and 904 are in the mid-IQ strata. There are 10,332
users with at least 10 edits, of which 176 are high-IQ and 1,116 are mid-IQ.
The median user with at least 10 annotations made their earliest annotation
in July 2013, so these accounts are sufficiently old to study behavior
over time in relation to (eventual) IQ.

In agreement with Fig.~\ref{AnnotTimeStat} and the IQ diamond model in
general, users with the highest IQ tend to make relatively more first
annotations on songs (Fig.~\ref{Lifespans}, top left). High-IQ users also make
relatively more first edits on annotations early in their lifespan, although
there is little difference with the general user population later in the
lifespan (Fig.~\ref{Lifespans}, top right). Thus, the positive relationship
between IQ and edit time rank (Fig.~\ref{EditTimeStat}, top left) is not likely caused by
high-IQ users making relatively fewer early edits, but by later edits more
likely coming from high-IQ users.

Quality of annotations in terms of quality tags and annotation length generally
increases over a user's lifespan (Fig.~\ref{Lifespans}, middle row). Users with high-IQ
use relatively more quality tags, especially early on;
however, annotation length from high-IQ users is comparable to those of
mid-IQ users. Again, annotation length is nuanced. Annotation
length has a mostly positive relationship with proportional time rank
(Fig.~\ref{AnnotTimeStat}, bottom right), and the discrepancy may come from
high-IQ users that make more first-annotations on songs
(Fig.~\ref{Lifespans}, top left). As hypothesized above, there may be time
pressure to annotate faster, driving down annotation length. Still,
mid- and high-IQ users create substantially longer annotations than other
users on the site, so there is some notion of quality or expertise
that is marked by annotation length.
Finally, high-IQ users tend to annotate relatively more original
lyric segments and songs (Fig.~\ref{Lifespans}, bottom row), which may
speak to their higher knowledge about lyrics as suggested by the IQ diamond model.

As the above information is informative for early detection of expertise, we
reiterate the properties of early contributions of
(eventual) high-IQ users on the site that distinguish them from other users. From the start,
high-IQ users make a higher proportion of first annotations and edits, use more quality tags
in their annotations, and annotate
more original lyric segments and songs. These traits are likely beneficial to the Genius
platform, and they appear to be inherent traits of high-IQ users upon entry to the site.


\section{Early Prediction of Super Experts}\label{sec:prediction}

\begin{table}[t]
  \begin{center}
    \caption{Predictors for classifying super experts and normal experts,
    derived from the analysis in Sections~\ref{sec:annotation_dynamics} and \ref{sec:evolution}.}
    \label{tab:predictors}
    \begin{tabular}{r l}
      \toprule
      $\alpha_1$ & mean \# of quality tags in first 15 annotations \\
      $\alpha_2$ & mean time between first 15 annotations \\
      $\alpha_3$ & \# of first 15 annotations that are a song's first \\
      $\alpha_4$ & mean originality on songs for first 15 annotations \\
      $e_1$ & mean time between first 15 edits \\
	  $e_2$ & \# of first 15 edits that are an annotation's first \\
      \bottomrule \\
    \end{tabular}

    \caption{Bootstrapped mean coefficients and confidence intervals of
      a logit model using the predictors in Table~\ref{tab:predictors}
      for the outcome variable of super-expert vs.\ normal expert.}
    \label{tab:logit}
    \begin{tabular}{r c c c c c c c}
      \toprule
      Predictor & mean regression coeff. & 95\% CI \\
      \midrule
      $\alpha_1$ & 0.8177 & (0.590, 1.072) \\
      $\alpha_2$ & 0.7942 & (0.548, 1.074) \\
      $\alpha_3$ & 0.5166 & (0.349, 0.700) \\
      $\alpha_4$ & 0.2623 & (0.098, 0.435) \\
      $e_1$ & -0.3832 & (-0.586, -0.193) \\
      $e_2$ & 0.2281 & (0.054, 0.409) \\
      \midrule
      intercept & 0.1127 & (-0.053, 0.283) \\
      \bottomrule \\
    \end{tabular}

    \caption{Classification results for various feature subsets.
      Listed results are the mean and standard deviation over 1,000 random 75\%/25\%
      training/test splits. Guessing the most common test label has
      mean accuracy of 0.522.}
    \label{tab:Incremental}
    \begin{tabular}{l c c}
      \toprule
      Predictors & Accuracy & AUC \\
      \midrule
      $\alpha_1$, $\alpha_2$, $\alpha_3$, $\alpha_4$, $e_1$, $e_2$ & $.673 \pm .029$ & $.748 \pm .030$ \\
      $\alpha_1$, $\alpha_2$, $\alpha_3$, $\alpha_4$, $e_1$        & $.671 \pm .029$ & $.744 \pm .030$ \\
      $\alpha_1$, $\alpha_2$, $\alpha_3$, $\alpha_4$               & $.659 \pm .030$ & $.733 \pm .031$ \\
      $\alpha_1$, $\alpha_2$, $\alpha_3$                      & $.659 \pm .028$ & $.727 \pm .031$ \\
      $\alpha_1$, $\alpha_2$                             & $.652 \pm .030$ & $.715 \pm .033$ \\
      $\alpha_1$                                    & $.616 \pm .032$ & $.674 \pm .034$ \\
      \bottomrule
    \end{tabular}
  \end{center}
\end{table}

Given the user behaviors discussed in earlier sections, we now
turn to early user expertise prediction. To do this, we continue
to use IQ as a proxy for expertise, as it is one simple metric that measures
contributions of all types on the site, and we set up a classification problem
that uses features derived from the first few annotations and edits of users.

First, we collected all users with at least 30 annotations and at least 30
edits. Of these users, we label the users in the highest third of IQ as ``super
experts,'' as these users are in the 99.8 quantile in IQ over all users on
Genius at the time of data collection. We label the lowest third of IQ as ``normal experts,'' as these users
are still in the 93.7 quantile of IQ (having at least 30 annotations and edits
leads to some accumulation of IQ). In total, we have 784 labeled users. Over
these users, the mean number of annotations per user is 109 and the mean number of edits per user is 537.

We use this coarse prediction framework for
several reasons. First, IQ is only a rough measurement of expertise. Second,
our IQ data was scraped at different times, and we do not have data on the
evolution of IQ over time for users. Nonetheless, since the distribution of
contributions to the site is heavy-tailed (Fig.~\ref{AnnEditDistr}), the most
active users contribute a large amount of content to the site, and we expect
that our IQ splits are still meaningful; super experts have over three
times higher mean number of annotations made than normal experts.
Early predictions of super experts can be beneficial to a site operator.
For instance, one can encourage these users with high potential to
continue to contribute to the site, or to contribute content
that requires expertise, such as annotations on lyrics of high originality~\cite{anderson2013steering,Zhang07}.

Next, we analyze a logistic model for predicting super vs.\ normal experts
from several content-based and edit-based predictors derived from our observations
of user behavior in the prior sections, such as the fact that experts use more
quality tags, work on more original songs, and often make first
edits (Table~\ref{tab:predictors}).
These predictors are computed over the first 15 annotations
and first 15 edits of the labeled users, and
were normalized to have zero mean and unit variance.
We use a bootstrapped logistic model with 10,000 samples
to estimate the mean and 95\% confidence intervals of the regression
coefficients (Table~\ref{tab:logit}).

The positive coefficients on $\alpha_1$, $\alpha_3$, $\alpha_4,$ and $e_2$ agree with our findings
in Section~\ref{sec:evolution} that the number of quality tags, proportion
of first annotations, originality of annotated songs, and proportion of first
edits, respectively, are higher early in the lifespan of high-IQ users. These
results also substantiate our IQ diamond model, which agrees with the importance
of first annotations in distinguishing expertise.

The fact that $\alpha_2$ (time between annotations) has a positive coefficient while
$e_1$ (time between edits) has a negative coefficient is surprising.  Users with
more time between their early annotations may face less time-pressure and may
make higher quality annotations.  On the other hand, users with less time
between their edits may simply make more edits.  If a user is making edits early
in their lifespan, then they may have an eye for good contributions, which
presumably provides a strong signal of expertise.

Finally, we evaluate these predictors in terms of test predictions.
To do so, we randomly split the data into 75\% for training and 25\% for test
and averaged the test accuracy and AUC over 1,000 splits (Table~\ref{tab:Incremental}).
Over the splits, this classifier attains a mean accuracy of 0.673 and mean AUC of 0.748,
whereas a majority-class baseline guess yields mean accuracy of 0.522.
This substantial performance gain is remarkable given that we only
use limited information from the first 15 annotations and edits.

\section{Discussion}

Crowdsourced information platforms provide a variety of information for the
world. Here, we have analyzed various aspects of the temporal dynamics of
content and users on the Genius platform that collects and manages crowdsourced
musical knowledge. Genius has a substantially different knowledge curation
process compared to Question-and-Answer platforms such as Stack Overflow or
formal authoritative sources like Wikipedia. In turn, we found that the content
and user dynamics have markedly different behavior from these other well-studied
platforms.

We modeled one new type of dynamics with an IQ diamond ($\Diamond$) model, which captures the
fact that eventual high-IQ users tend to annotate songs early and late on, with
low-IQ users annotating in between. Even though the IQ diamond model is
just one possible explanation for this user behavior, the ideas could
potentially be used for mechanism design on platforms governed by dynamics
similar to Genius. For example, to encourage annotation coverage of a certain
song, one could incentivize (eventual) experts to make just a few annotations
(e.g., by adjusting IQ incentives). Under the IQ diamond model, this will open
the song to the bulk of users to create more annotations. Also, if higher
quality annotations are desired, one may incentivize experts to edit
intermediate annotations appearing at the bottom of the ``U-shaped'' curve
(middle of the diamond). Furthermore, some users may desire to annotate a song
but do not have the skills themselves. In this case, if they annotate some less
original or ``easy'' lyrics, the induced network effects may incentivize
experienced users to create high quality annotations.

\begin{figure}[t]
	\centering
	\includegraphics[width=\columnwidth]{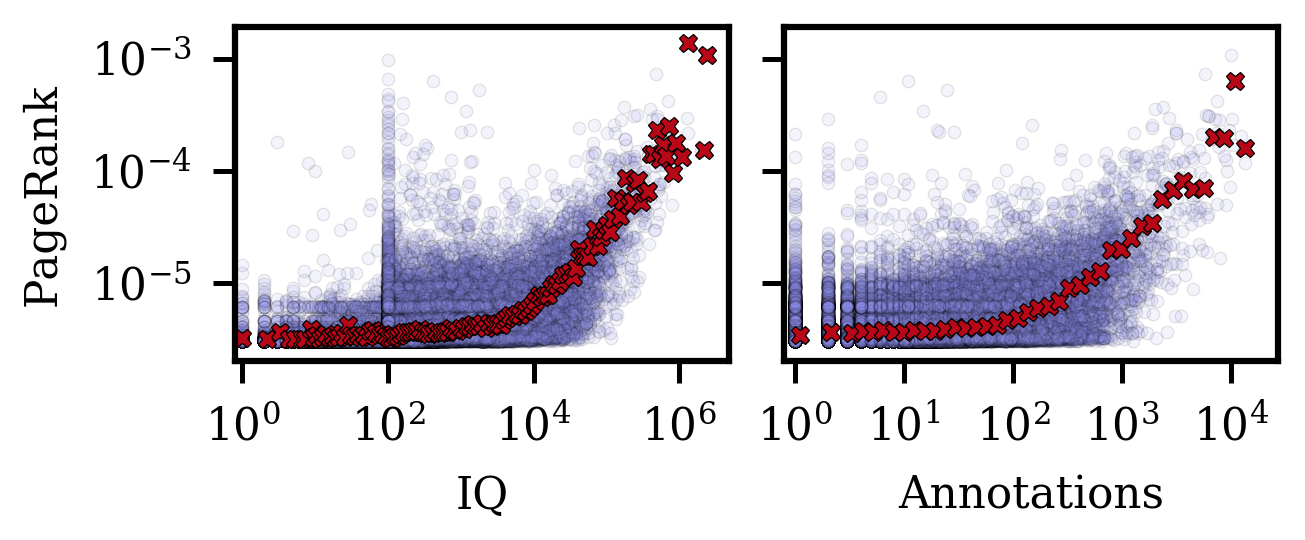}
	\caption{There is a positive relationship between user PageRank scores in the Genius
          social network and the IQ of the user (left) or number of annotations made by the user (right).
          Each blue point represents a user, and red X's are binned means.
          Many users have exactly 100 IQ, which is the amount awarded for adding a profile picture.}
	\label{PagerankStats}
	\vspace{-10pt}
\end{figure}

Studying user behavior over time stratified by eventual expertise revealed
several traits of eventual experts that can be identified by early behavior. We
used these findings to develop strong predictors of eventual experts. The set
of predictions could be enhanced by social features, as Genius has an underlying
social network for its users. We collected the social network data of 782,432
users and 1,777,351 directed ``following'' relationships among them, although we do not
have temporal information on the social network that would be useful for
early prediction. Still, we found that certain social features are strongly
correlated with user expertise. For instance, Fig.~\ref{PagerankStats} shows
that users with large PageRank scores have higher IQ and have made more
annotations. Just using PageRank as a predictor for our experiments in
Section~\ref{sec:prediction} achieves a mean AUC of 0.972, and even just using
the in-degree achieves a mean AUC of 0.977. While determining current expertise
is often a useful task for crowdsourced-information sites, it is not a difficult
task when we define expertise solely based on a public metric such as IQ.

\xhdr{Future directions}
Although contribution dynamics on Genius are different compared to platforms
such as Stack Overflow, Amazon, and Yelp, examining other types of sites
with similar structure is a promising direction for future research.
In particular, we hypothesize that 
platforms that have pages with separate but related open-ended contribution prompts
may have similar contribution dynamics to that of Genius. One example is the
Online Encyclopedia of Integer Sequences (\texttt{oeis.org}). The site has entries for
integer sequences appearing in mathematics, and there are several open-ended ways to contribute
to various structured ``fields'' that contribute to understanding the sequence
(e.g., formulas, code to compute the sequence, or references to papers that use the sequence).
Such fields are similar to lyric segments on Genius, and contributions on a sequence
might have similar dynamics to those of annotations on a song.
Also, there may be related user contribution on sites that have been well-studied.
For instance, on Stack Overflow, if there are multiple questions asked in quick succession about a specific topic,
then the arrival of accepted answers to these questions may follow similar dynamics
to the arrival of annotations to different lyric segments on a song on Genius.

There are also many avenues for future work based on this Genius data or the models
we have developed for Genius. For one, we could design experiments based on the
actionable mechanism designs described above. Also, one could use
the Genius data to augment other music datasets~\cite{Brost19,Bertin11}.
Conversely, other music data may improve the analysis of the content on
Genius. One may argue that Genius is somewhat limited by its focus on lyrics
since the musical context of the lyrics is deeply important for both the
analysis and experience of songs~\cite{Kehrer16}.
However, as described in Section~\ref{sec:diamond}, there are several
possible extensions or improvements that one could make with the IQ diamond model
that could be applicable outside of Genius.

Moreover, there is plenty of information on Genius that we did not collect or
analyze. In particular, ``verified'' annotations (annotations written by
artists) and forms of user contribution besides annotations and edits are
available. The rich linguistic information in the lyrics and annotations
can also be analyzed in more depth. For instance, we did not consider
the sequential and hierarchical structures in lyrics that have been used in
music information retrieval~\cite{Tsaptsinos17}. Such structure adds yet another
layer of depth to the organization of contributions and content on Genius that
distinguishes it from other crowdsourced information sites.

\section*{Acknowledgements}

This research was supported in part by NSF Award DMS-1830274, ARO Award W911NF19-1-0057, ARO MURI, and JPMorgan Chase \& Co.

\fontsize{9.0pt}{10.0pt} \selectfont 

\bibliography{shorterbib}
\bibliographystyle{aaai}

\end{document}